\documentclass[%
reprint,
amsmath,amssymb,
aps,
nobibnotes
pra,
]{revtex4-1}

\usepackage{graphicx}
\usepackage{dcolumn}
\usepackage{bm}
\usepackage[dvipsnames]{xcolor}

\usepackage{lineno}
\usepackage{gensymb}
\usepackage{amsmath}
\usepackage{mathtools}
\usepackage{braket}
\usepackage{comment}

\begin{document}

\preprint{APS}

\title{Quantum Imaging of Birefringent Samples using Hong-Ou-Mandel Interference}

\author{Carolina Gonçalves}
\author{Tiago D. Ferreira}
\email{tiago.d.ferreira@inesctec.pt}
\author{Catarina S. Monteiro}
\author{Nuno A. Silva}
\email{nuno.a.silva@inesctec.pt}

\address{Departamento de Física e Astronomia, Faculdade de Ciências, Universidade do Porto, Rua do Campo Alegre s/n, 4169-007 Porto, Portugal}
\address{INESC TEC, Centre of Applied Photonics, Rua do Campo Alegre 687, 4169-007 Porto, Portugal}

\date{\today}

\begin{abstract}
Two-photon interference in a Hong–Ou–Mandel (HOM) interferometer can be used as a quantum sensing mechanism due to the sensitivity of the interference dip to perturbations of the photon indistinguishability. In particular, recent works have generalized this concept to microscopy setups, but the sensitivity to optical path differences constrains its application to samples with thickness variation typically below a few micrometers if tracking changes in the coincidences at a fixed delay. Extending the concept to polarization microscopy and circumventing this limitation, this manuscript explores the use of a narrowband photon pair source with coherence length $>1$ mm to broaden the HOM dip. Thus, realistic sample-thickness variations introduce negligible temporal distinguishability, and changes in coincidence rate at the dip centre are then dominated by sample-induced polarization effects. To compute the polarization rotation, we develop a statistical model for the interferometer, derive the Fisher information, and establish a maximum-likelihood estimator for the local fast-axis angle. Recording dip and baseline frames at each sample position via raster scanning, the experimental results validate the framework, agreeing with classical polarized-intensity images while demonstrating operation at a maximum-precision regime and insensitiveness to layer thickness. Overall, the approach enclosed provides a quantum-based quantitative imaging of birefringent structures, which can motivate further advantageous applications, including enhanced signal-to-noise ratio and lower damage imaging of photosensitive samples.
 
\end{abstract}

\maketitle

\section{Introduction}

Polarization imaging and microscopy is an interesting technique that can be used to recover information that is invisible to the naked eye by measuring spatially resolved Stokes or Mueller matrices and inferring properties such as anisotropy, diattenuation, depolarization, chiral response, and stress distributions \cite{parkinson2024mueller}. This information supports applications in diverse subjects from biomedical diagnosis \cite{he2021polarisation} to remote sensing \cite{tyo2006review} and material characterization \cite{xie2025quantum_GRA}, just to mention a few examples. Classically, implementations range from Mueller-matrix polarimetry and quantitative polarized-phase microscopy \cite{dragomir2007quantitative} to polarization-sensitive optical coherence tomography \cite{de2017polarization}. However, because these schemes rely on classical light and intensity-based detection, they are fundamentally limited by shot noise, reduced sensitivity under low-photon-flux conditions, and trade-offs between spatial resolution, acquisition time, and sample damage, particularly for the characterization of weakly scattering or photosensitive samples \cite{moreau2019imaging}. In recent years, these limitations have motivated a growing interest in the integration of quantum light sources into imaging systems \cite{moreau2019imaging}, promising to surpass classical limits with enhanced measurement accuracy, particularly under low-light conditions \cite{xie2025quantum_GRA}, and nonlocal capabilities with significant advantages for remote sensing applications \cite{besaga2024nonlocal,pedram2024nonlocality}.

From a broad perspective, quantum polarimetric sensing techniques can be categorized into fixed-delay correlation-based schemes and interferometric approaches exploiting Hong–Ou–Mandel (HOM) two-photon interference. In fixed-delay correlation schemes, experimental implementations typically employ polarization-entangled photon pairs in a polarizer–sample-analyzer configuration to characterize both the phase retardance and optical-axis orientation of anisotropic materials. The birefringence parameters are inferred from coincidence photon counts following well-defined polarization transformations applied across the sample or reference arm. This approach enables nanometer-scale precision in phase-retardance estimation \cite{lyons2018attosecond} and angular accuracies below $1^{\circ}$ for the optical-axis orientation \cite{harnchaiwat2020tracking,sgobba2023optimal}, while exhibiting enhanced robustness in low-photon-flux and noisy environments due to the intrinsic noise rejection afforded by coincidence detection \cite{xie2025quantum_GRA,zhang2024quantum}. Fixed-delay correlation schemes encompass a range of related protocols, including quantum optical rotatory dispersion \cite{tischler2016quantum}, ghost polarimetry \cite{restuccia2022measuring}, and distributed or nonlocal polarimetric measurements based on polarization-entangled photon pairs. In particular, the latter exploits polarization transformations exclusively in the reference arm, allowing nonlocal extraction of the polarization properties of the sample, thus minimizing optical complexity and perturbation in the measurement arm and enabling remote polarimetric sensing modalities \cite{xie2025quantum_GRA,besaga2024nonlocal,pedram2024nonlocality}.

On the contrary, HOM–based schemes explore two-photon interference to infer birefringence from changes in the depth and shape of the coincidence dip as a function of reference delay and relative polarization alignment of the photons in the two interferometer arms. With the polarization in the control arm typically kept fixed \cite{sukharenko2024polarization_HOM,yung2023jones,yung2022polarization}, depth-resolved birefringence profiles can be obtained with techniques such as quantum optical coherence tomography (QOCT) and its polarization-sensitive variants (PS-QOCT), being immune to even-order dispersion \cite{abouraddy2002quantum, booth2004polarization,booth2011polarization}. Yet, to increase sensitivity, most HOM-based schemes utilize ultrabroadband photon-pair sources, yielding coherence lengths of only tens of micrometres \cite{ndagano2022quantum,booth2011polarization}. This is ideal for depth resolution, but makes the HOM dip highly sensitive to small changes in sample thickness, strongly limiting the application of the technique. 

To circumvent this limitation, this work explores the use of a narrowband single-photon source with a coherence length exceeding $1$~mm to deliberately broaden the HOM dip. Operating at the broad, flat minimum of the dip, the measurement is intrinsically insensitive to small depth variations, making the coincidence rate governed by polarization-induced distinguishability rather than path-length fluctuations and enabling accurate characterization of the intrinsic birefringence of samples. Leveraging on this concept, we describe the implementation of a two-dimensional raster-scan quantum birefringence imaging technique, providing a theoretical framework of a statistical estimator for the birefringence angle that incorporates losses \cite{lyons2018attosecond} that is shown to maximize the Fisher information and saturate the Cramér-Rao bound. 

\section{Methodology}
For sensing applications, Hong–Ou–Mandel (HOM) interference provides a sensitive probe of the environment by exploiting two-photon indistinguishability. Indeed, when two photons are combined on a balanced beam splitter, their coincidence rate of the output ports exhibits a characteristic dip whose depth is set by the overlap of the quantum states of the individual photons across all degrees of freedom (temporal, spectral, spatial, and polarization) \cite{hong1987measurement,bouchard2020two,harnchaiwat2020tracking,sgobba2023optimal,altuzarra2019imaging,yung2023jones,yung2022polarization,yoon2024quantum,chen2019hong}.
In the particular case of polarization sensing, the introduction of a birefringent sample in one of the paths induces distinguishability between the polarization states of the photons, which reduces the quantum interference visibility and decreases the depth of the HOM dip up to its disappearance for orthogonal polarization states. 
The core of this work is set on the implementation of a microscopy setup that performs a raster scan of a sample and monitors the depth of the HOM interference dip, correlating it with the polarization state change and birefringence of the sample. Yet, contrary to other approaches previously reported in the literature\cite{booth2011polarization, sukharenko2024polarization_HOM}, our solution explores a narrowband single-photon source to suppress sensitivity to variations in sample thickness, ensuring that the measured dip depth primarily reflects the birefringent properties of the imaged sample.

\subsection{Theoretical Model}

\begin{figure}[ht!]
\centering\includegraphics[width=0.5\textwidth]{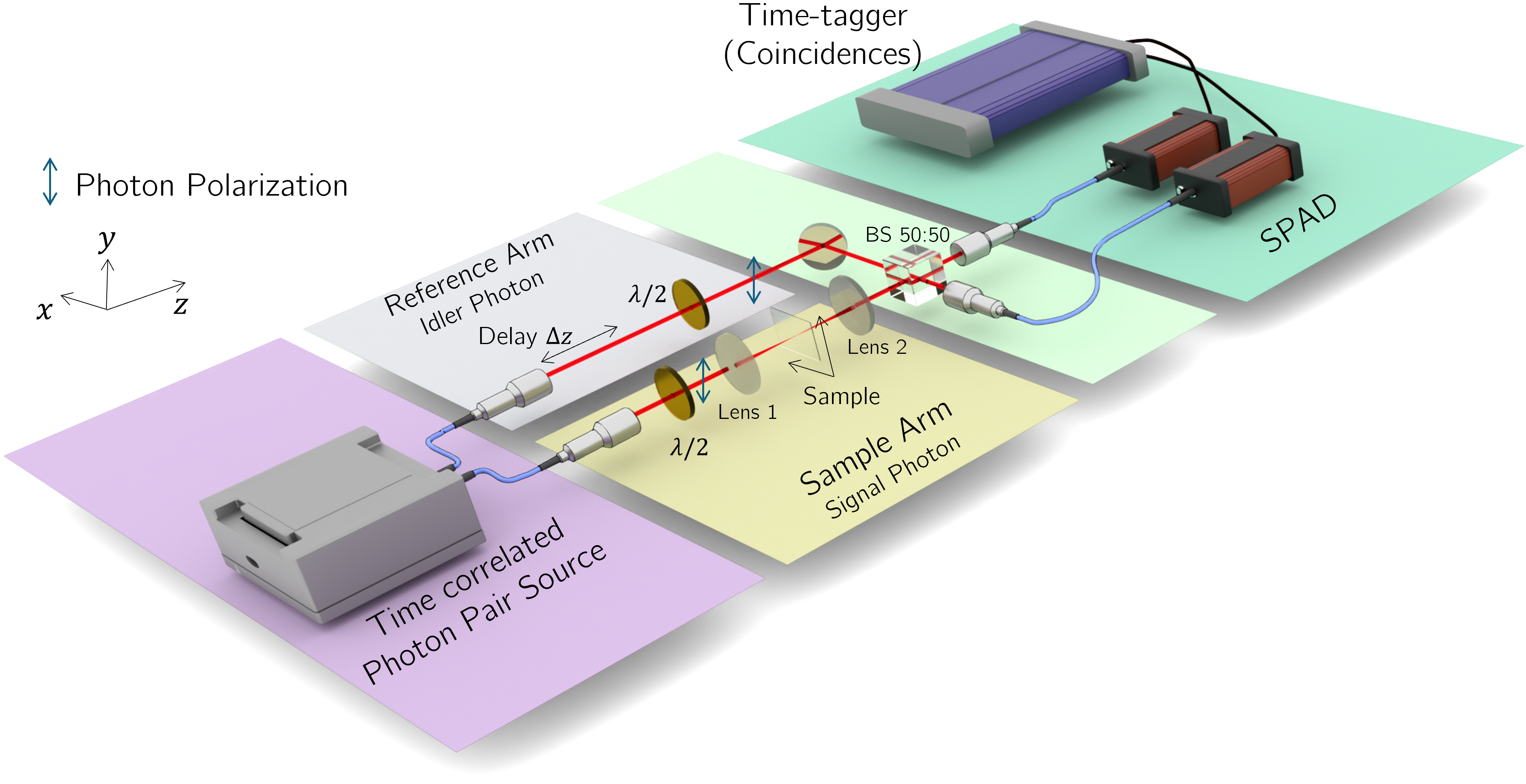}
\caption{Schematic of the experimental setup. Photon pairs from a commercial narrowband SPDC source are separated into reference and sample arms. Polarizing beam splitters (not shown) and half-wave plates prepare and control the polarization in each arm. In the sample arm, two $f=50\,\mathrm{mm}$ lenses focus the photons onto the sample, which is mounted on motorized $x$--$y$ translation stages for raster scanning. The reference arm includes a motorized translation stage that controls the relative delay $\Delta z$, thereby setting the HOM-dip position. The photons are then recombined at a $50:50$ beam splitter and detected with single-photon avalanche diodes. The coincidence events are recorded with a time-tagger. Bandpass filters centered at $810\,\mathrm{nm}$ (not shown) suppress stray light.}
\label{fig:img-setup}
\end{figure}

To set a theoretical framework for our HOM polarization microscopy setup, conceptually depicted in Figure \ref{fig:img-setup}, we start by assuming that immediately before the beam splitter, the two-photon polarization state can be generically described as
\begin{equation}
\ket{\psi_0}=\frac{1}{\sqrt{2}}\left(
\ket{\vec{\phi}_1}_S\ket{\vec{\phi}_2}_I+
\ket{\vec{\phi}_2}_S\ket{\vec{\phi}_1}_I\right).
\label{eq:input_state}
\end{equation}
Here the subscript labels the spatial path, so $\ket{\vec{\phi}_1}_{S,I}$ denotes a signal ($S$) / idler ($I$) photon with polarization $\vec{\phi}_1$. Before the sample, both photons are set to horizontal polarization, $\vec{\phi}_1=\vec{\phi}_2$, using half-wave plates and polarizing beam splitters. 

To model the effect of the sample in the HOM dip, we start by considering the effect of the polarization rotation. For simplicity, we assume that the sample in the signal arm is a linear retarder characterized by its retardance $\delta$ (phase delay between orthogonal principal axes) and fast-axis orientation $\theta$ measured from the laboratory horizontal \cite{booth2004polarization}. In the lab basis, the sample Jones matrix is
\begin{equation}
J(\theta) = R(-\theta)
\begin{pmatrix}
e^{i\delta/2} & 0 \\
0 & e^{-i\delta/2}
\end{pmatrix}
R(\theta),
\end{equation}
where $R(\theta)$ is the rotation matrix. The complete operator becomes
\begin{equation}
    \hat{J}_T=\hat{J}_S(\theta)\otimes\hat{I}_I
\end{equation}
with $\hat{I}_I$ corresponding to the identity operator for the reference photon, introducing polarization distinguishability between the interferometer arms.

In a typical HOM configuration, both photons are then combined using a beam splitter, where the total operator can be constructed as
\begin{equation}
    \hat{U}_{BS}=\hat{U}_{BS}^S\otimes\hat{U}^I_{BS}=\begin{pmatrix}
t & r \\
r & t
\end{pmatrix}\otimes\begin{pmatrix}
t & r \\
r & t
\end{pmatrix}
\end{equation}
where $t$ and $r$ are the transmission and reflection coefficients, respectively. The final state after reaching the detectors can then be computed using
\begin{equation}
    \ket{\psi_f}= \hat{U}_{BS}\hat{J}_T\ket{\psi_0}.
\end{equation}

However, besides the polarization changes, the presence of the sample also introduces an optical path delay on the signal photon. To model this effect, we account for the time delay introduced by optical path differences by describing the temporal wavepacket as a Gaussian in the path-length coordinate $z$ (related to delay via $z=c t$) as $f_i(z)=(8/\pi l_c^4)^{1/4}\exp[-2(z-z_i)^2/l_c^{2}]$, where $z_i$ denotes the center of the wavepacket at $t=0$. The wavepacket is normalized so that $\int |f_i(z)|^{2} dz = 1$. As the delay difference $\Delta z = z_2 - z_1$ departs from the dip center, photon indistinguishability is progressively reduced which can be more accurately modeled by describing the two-photon state as a mixture state \cite{dibrita2023easier}: 
\begin{equation}
  \rho(\Delta z)=p(\Delta z)\,\rho_{\mathrm{indist}}+\bigl[1-p(\Delta z)\bigr]\rho_{\mathrm{dist}},
  \label{eq:rho_mixture}
\end{equation}
where $\rho_{\mathrm{indist}}$ and $\rho_{\mathrm{dist}}$ represent the perfectly indistinguishable and fully distinguishable limits of the density matrix, respectively. For Gaussian wavepackets, the probability that the photons are indistinguishable is given by the overlap integral $p(\Delta z)=\int_{-\infty}^{\infty}f_1(z)f_2(z)dz=\exp\bigl[-\Delta z^{2}/l_{c}^{2}\bigr]$. Thus, ~\eqref{eq:rho_mixture} provides a unified description of the two operating regimes of the setup. When $\Delta z \approx0$, the photons are indistinguishable and the system is described by $\rho_{\mathrm{indist}}$, corresponding to the center of the HOM dip. In contrast, for $|\Delta z|\gg l_{c}$, the photons become fully distinguishable and the state approaches $\rho_{\mathrm{dist}}$, yielding the baseline coincidence level.

Finally, in the HOM interferometer we are interested in the coincidence probability, $P_c$, which is the probability of measuring a single photon in both detectors within the coincidence window. For the current model, this probability is calculated through the expected value of the coincidence operator $\hat{P}_c$ as $P_c(\Delta z,\theta)=\mathrm{Tr}\left(\rho(\Delta z)\hat{P}_c\right)$. Following \cite{dibrita2023easier}, the coincidence probability is a (generally cumbersome) function of $\theta$ and $\delta$. To validate the methodology for the purpose of this work, we employed a synthetic sample composed of polymer retarder film fragments, enabling a controlled and simplified analysis.
Although the film is not a strict half-wave retarder at $810~nm$, as the design wavelength is at $560~nm$, it provides a spatially structured birefringent target, and the visibility contrast remains governed by polarization-induced distinguishability. This case simplifies the mathematical model by assuming that $\delta\approx\pi$. Further, assuming a 50:50 beam splitter, $t=1/\sqrt{2}$ and $r=i/\sqrt{2}$, we get a final expression for the coincidence probability as
\begin{equation}
P_c(\Delta z,\theta)=\frac{1}{2}\bigl[1-\alpha e^{-\Delta z^{2}/l_c^{2}}\cos^{2}2\theta\bigr],
\label{eq:coincidence_probability}
\end{equation}
where $\alpha\le1$ quantifies the maximum visibility obtained experimentally \cite{lyons2018attosecond}. It is straightforward to conclude that in the dip position, assuming $\Delta z\approx0$, the visibility changes are dominated by the polarization according to a cosine term, resembling the classical Malus' law. Also note that for our Gaussian temporal model, the indistinguishability factor is associated with a term $e^{-\Delta z^{2}/l_c^{2}}$, which for a change of $1\%$ would require a path delay of $\Delta z\approx100/(n_{sample} - 1)$ $\mu$m, assuming $l_c=1$ mm. Thus, if the delay is negligible compared to $l_c$ (which is the case of a narrowband single-photon source, $l_c\approx1mm$), the overlap integral $p(\Delta z)$ can be considered constant across all points of the sample.

\subsection{Statistical Analysis and Polarization Angle Estimator}
Although we already obtained a formal expression for the change of $P_c$ with $\theta$, we still need a statistical analysis to determine what is the optimal method for extracting the polarization rotation angle $\theta$ directly from the raw photon-count records and that accounts the presence of experimental noise and other imperfections. For this we follow a standard multinomial estimation theory, as discussed in \cite{lyons2018attosecond}. We start by assuming that the HOM interferometer has three mutually exclusive outcomes: no clicks on either detector ($i=0$), a click on one detector ($i=1$), and simultaneous clicks on both detectors ($i=2$). Introducing the probability of one photon being lost as $\gamma$, the probabilities of each of the outcomes can be expressed as a function of the coincidence probability $P_c(\Delta z,\theta)$ as
\begin{equation}
    \begin{pmatrix}  
P(0|\theta)\\  
P(1|\theta) \\
P(2|\theta)
\end{pmatrix} = \begin{pmatrix}  
\gamma^2 & \gamma^2 \\  
2\gamma(1-\gamma) & 1-\gamma^2\\
1 - \gamma^2 - 2\gamma(1-\gamma) & 0
\end{pmatrix}\begin{pmatrix}  
P_c(\Delta z,\theta)\\  
1 - P_c(\Delta z,\theta)
\end{pmatrix},
\end{equation}
where $1-P_c(\Delta z,\theta)$ is interpreted as the probability of measuring a bunching event. The probability $P(i|\theta)$ corresponds to the probability of measuring each outcome $i$ giving that the sample introduces a rotation of $\theta$ on the polarization of the photons of the sample arm.

\subsubsection{Fisher Information}
In order to proceed with a statistical theory for the quantum measurement, it is useful to look first at the Fisher information(FI) to determine the conditions of optimal operation and minimize the free parameters. In generic terms, FI quantifies how much information the experimental data have about the parameter that we want to measure, in this situation $\theta$. For the statistical model presented above, the FI can be computed as
\begin{equation}
\begin{split}
&F_{\theta}=\sum_i\frac{\left(\partial_{\theta} P(i|\theta)\right)^2}{P(i|\theta)} \\
=&\frac{16 \alpha^{2} \left(\gamma - 1\right)^{2} \left(\gamma + 1\right) \sin^{2}{\left(2 \theta \right)} \cos^{2}{\left(2 \theta \right)}}{\left(\alpha \cos^{2}{\left(2 \theta \right)} - e^{\frac{\Delta z^{2}}{l_c^{2}}}\right) \left(\alpha \cos^{2}{\left(2 \theta \right)}(\gamma - 1) - e^{\frac{\Delta z^{2}}{l_c^{2}}}(3 \gamma  + 1)\right)}.
\end{split}
\label{eq:fisher_expression}
\end{equation}
The FI is directly linked to the Cramér–Rao inequality \cite{Kok_Lovett_2010}, which establishes a lower bound on the variance ${\sigma}^2_\theta$ of any unbiased estimator as $Var\left(\theta\right) \geqslant \left(NF_\theta\right)^{-1}$. Thus, for a given $\theta$, it can be verified numerically that the FI is maximized for $\Delta z=0$ for all relevant $\theta$ and $\gamma$ values, consistent with the intuitive expectation that the dip centre provides maximum sensitivity to visibility changes. Note, however, that the FI is not constant for all $\theta$ values, Figure \ref{fig:estimation_fisher_cramer}, being minimal for angles multiple of $\pi/4$, which corresponds to positions of large uncertainty in the angle estimation.

\subsubsection{Maximum-Likelihood Estimator and Polarization image}
The maximum-likelihood estimator (MLE) for $\theta$, after $N=N_{0}+N_{1}+N_{2}$ detection events, follows a multinomial distribution and is given by
\begin{equation}
L(\theta)\;=\;\Pr(\mathbf N\!\mid\!\theta)
           =\frac{N!}{N_{0}!\,N_{1}!\,N_{2}!}\,
             \prod_{i=0}^{2}P(i|\theta)^{N_{i}}.
\label{eq:multinomial_likelihood}
\end{equation}
Thus, the estimator that maximizes the determination of $\theta$ is obtained by extremizing $L(\theta)$, or equivalently its logarithm $\mathcal{L}(\theta)=\sum_{i}N_{i}\ln P(i|\theta)$. For a single parameter, this corresponds to finding the value of $\theta$ that satisfies the stationary condition $\partial\mathcal{L}(\theta)/\partial\theta = 0$ at the dip position $\Delta z=0$, the point of maximum information, with the constraint $N_0+N_1+N_2=N$. Solving for this condition yields
\begin{equation}
    \alpha\cos^22\tilde{\theta}=\underbrace{\frac{N_{1} \gamma - N_{1} + 3 N_{2} \gamma + N_{2} }{ (\gamma - 1)(N_{1} + N_{2})}}_{V},
    \label{eq:angle_equation}
\end{equation}
where the right-hand side can be shown to correspond to the visibility $V$ of the HOM interferometer expressed as a function of the counts $N_1$ and $N_2$, and the photon-loss parameter $\gamma$. The $\tilde{\theta}$ is then retrieved using the procedure presented by Harnchaiwat et al. in Ref. \cite{harnchaiwat2020tracking},

\begin{equation}
\theta=
\begin{cases}
\tilde{\theta}, & \text{if } N_1 - N_2\left(\dfrac{1+3\gamma}{1-\gamma}\right)\ge 0,\\[6pt]
\pm\dfrac{\pi}{2}, & \text{if } N_1 - N_2\left(\dfrac{1+3\gamma}{1-\gamma}\right)< 0,
\end{cases}
\end{equation}
\begin{equation}
\theta=
\begin{cases}
\tilde{\theta}, &
\text{if }\sqrt{\dfrac{N_1 - N_2\left(\dfrac{1+3\gamma}{1-\gamma}\right)}{\alpha\left(N_1+N_2\right)}}\subset[-1,1],\\[10pt]
0, &
\text{if }\sqrt{\dfrac{N_1 - N_2\left(\dfrac{1+3\gamma}{1-\gamma}\right)}{\alpha\left(N_1+N_2\right)}}\not\subset[-1,1].
\end{cases}
\end{equation}
where $\theta$ is the measured angle constrained to values that make equation \ref{eq:angle_equation} true. Note that since $V\propto \cos^2 2\theta$, a single-visibility measurement cannot distinguish between $\theta$ and $\pi/2-\theta$, leading to a two-fold ambiguity. Resolving this requires at least two reference polarization configurations (e.g. horizontal and $45^\circ$), which we leave out of the scope of this work.

Finally, to compose an image of the polarization rotation angle (associated with the birefringence), we raster-scanned the sample in the transverse plane, and for each pixel $(m,n)$ two images were acquired: a dip frame at $\Delta z\approx0$ and a baseline frame at $|\Delta z|\gg l_c$. The corresponding $N_i$ counts, were recorded for equal acquisition times and combined to generate a normalized visibility map
\begin{equation}
\begin{split}
V(\theta)_{(m,n)} = \left.\frac{N_{1}^{(m,n)} \tilde{\gamma} - N_{1}^{(m,n)} + 3 N_{2}^{(m,n)}  \tilde{\gamma} + N_{2}^{(m,n)} }{ ( \tilde{\gamma} - 1)(N_{1}^{(m,n)} + N_{2}^{(m,n)})}\right|_{\Delta z=0},
\end{split}
\label{eq:visibility_map}
\end{equation}
where the photon loss probability $\tilde{\gamma}$ was calibrated far from the dip
\begin{equation}
     \tilde{\gamma}=\gamma_{(m,n)}=\left.\frac{N_{1}^{(m,n)}-N_{2}^{(m,n)}}{N_{1}^{(m,n)}+3N_{2}^{(m,n)}}\right|_{|\Delta z|\gg l_c}
     \label{eq:gamma}
\end{equation}
which we allow to be position-dependent due to spatially varying absorption or scattering in the sample. The motivation for this is that different regions of the sample can exhibit different absorption levels, which produces coincidence rates that are dependent on the sample position. By acquiring the $\gamma$ for each sample position, we effectively reduce this absorption-related bias, leading to a more accurate estimation of the birefringence parameter $\theta$.

\subsection{Experimental Results}

For the experimental work we implemented a typical HOM setup conceptually depicted in Figure \ref{fig:img-setup}. Photon pairs were generated by a commercial source (Thorlabs SPDC810N) via spontaneous parametric down-conversion in a type-II periodically poled PPKTP crystal, pumped by a continuous-wave $405$~nm diode laser. The crystal temperature was actively stabilized at $28.20\pm0.05^{\circ}\mathrm{C}$ to ensure frequency degeneracy, producing a spectral bandwidth $\Delta\lambda<0.25$~nm and a single-photon coherence length $l_c\simeq\lambda^{2}/\Delta\lambda$~mm. This yields a HOM dip with a $1/e$ half-width exceeding $\sim1$~mm. The down-converted photons were separated into a signal arm, which contained the sample, and an idler arm serving as a reference. A motorized translation stage in the reference arm introduced a variable path delay $\Delta z$ before the two beams were recombined at a balanced, non-polarizing $50{:}50$ beam splitter. In the sample arm, a lens system (focal length $L=50$~mm) focused the photons onto the sample to increase the spatial resolution of the mapping. The probing position is controlled by motorized stages in the $X$ and $Y$ axis. Coincidences at the two output ports were recorded using single-photon avalanche diodes (SPADs) connected to a time-correlated single-photon counting module. In this work we used the Thorlabs SPDMH2F, with $60\%$ efficiency at $810$~nm, a dead time of $45$~ns, and a maximum dark count rate of $100$~Hz. All results reported here use a $1$~ns coincidence window.

\begin{figure}[ht!]
\centering\includegraphics[width=0.45\textwidth]{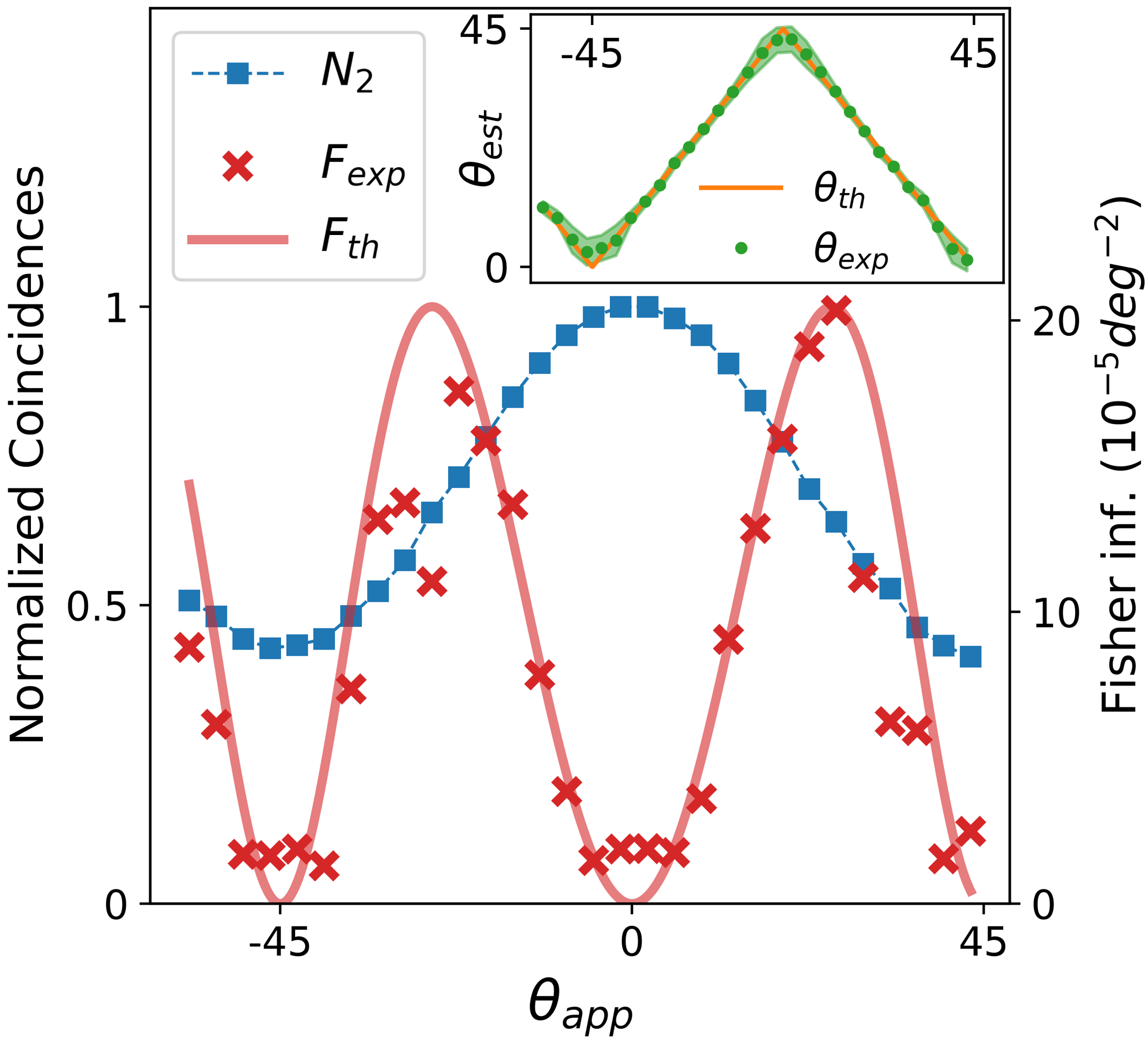}
\caption{Normalized coincidences as a function of the sample photon polarization angle measured at $\Delta z\approx0$, left axis. On the right axis we have the inverse variance $1/(NVar(\theta_{exp}))$ obtained from $N\sim3\times10^7$ repeated acquisitions at each angle (right axis), compared with $F_{\theta}$ from equation \ref{eq:fisher_expression}. The inset shows the corresponding angle estimates obtained using Eq.~\ref{eq:angle_equation}. Because the input angles are known, we can readily resolve the angle degeneracy.}
\label{fig:estimation_fisher_cramer}
\end{figure}

We began by probing the HOM dip as a function of the sample-photon polarization angle to validate the statistical model developed for our experimental setup and to assess its ability to estimate the polarization angle rotation. For this, we utilized a $\lambda/2$ waveplate in the sample position, which applied a polarization angle rotation $\theta_{app}$ to the signal photon, controlled by a rotation stage. First, by computing the experimental Fisher information for each applied angle, the results of Figure \ref{fig:estimation_fisher_cramer} demonstrate that we are operating in the maximum-precision regime allowed by our setup, as evidenced by the saturation of the Cramér–Rao bound set by the theoretical expression. In the Figure \ref{fig:estimation_fisher_cramer} inset, we also demonstrate that we can accurately recover the polarization angles, with an error below $3\%$ with the expected loss of precision at $\theta=k\pi/4$. Indeed, it is visible that the precision of the angle determination is not the same for all angles, with reduced sensitivity and increased variance near $\theta=k\pi/4$, where $\partial_{\theta}\cos^22\theta\approx0$. Note that in this case, because we know the angles at which the sample-photon polarization is set, we can resolve the angle degeneracy and correctly estimate the angle over the full range.
\begin{figure}[ht!]
\centering\includegraphics[width=0.35\textwidth]{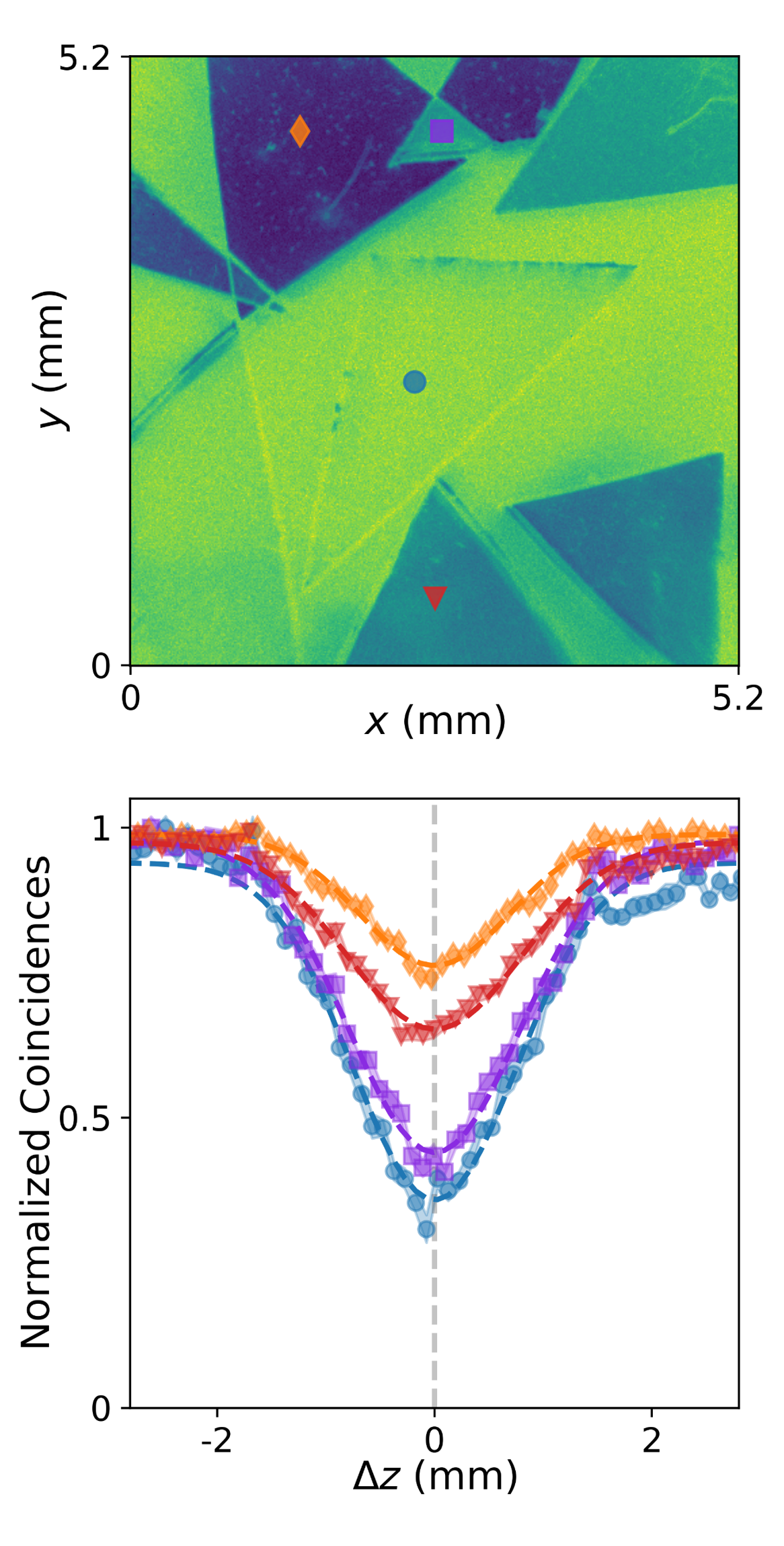}
\caption{HOM dip as a function of delay for different points on the sample. Top: false-color image of the sample acquired with a CMOS camera and a polarizer. Bottom: corresponding HOM dips as a function of delay, with markers indicating the sample positions at which each dip was measured. We measured each curve $10$ times with each data point acquired for $T=5.25$~s.}
\label{fig:HOM_vs_number_of_HWP}
\end{figure}

\begin{figure}[ht!]
\centering\includegraphics[width=0.4\textwidth]{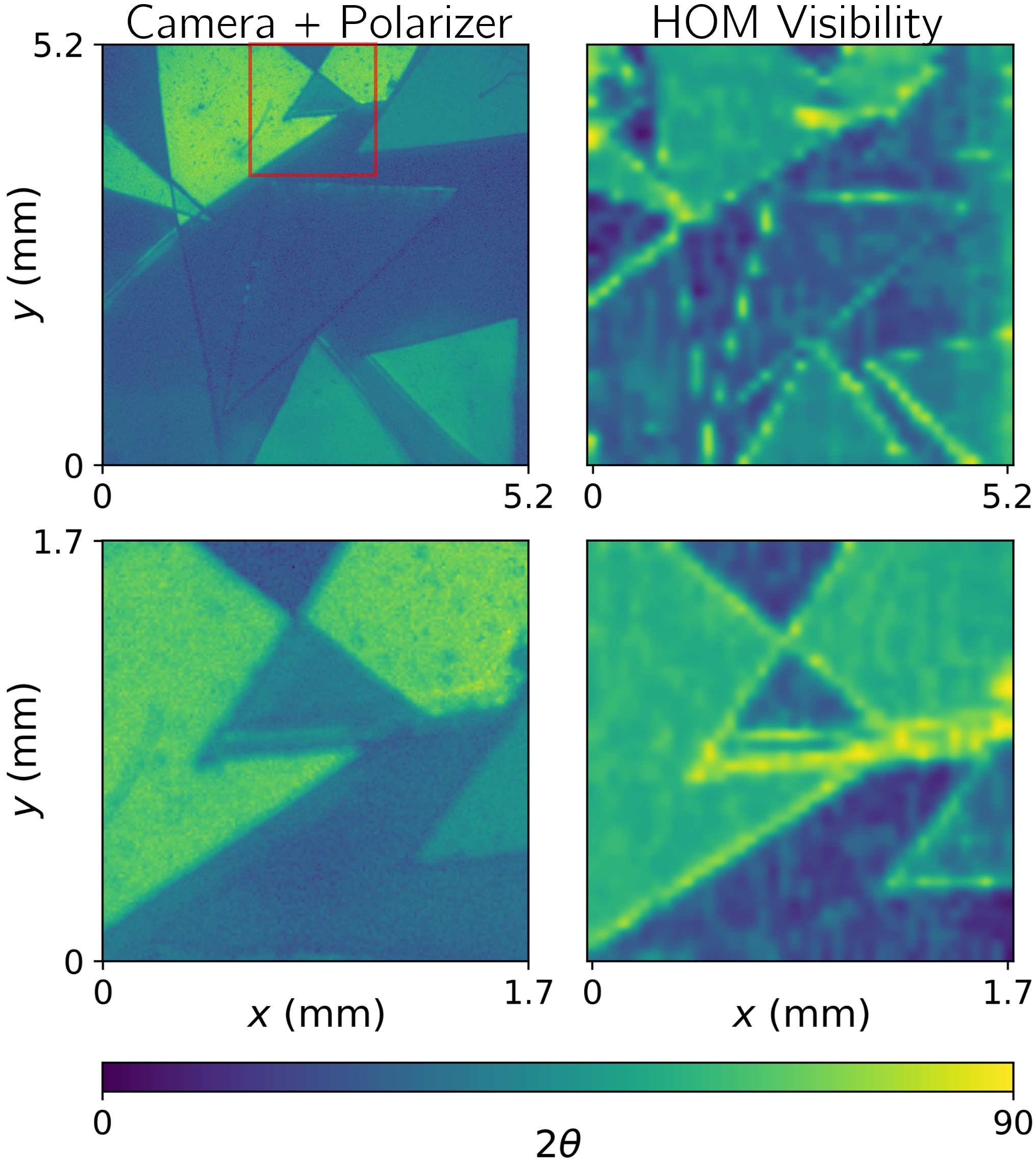}
\caption{Result of the spatial variation of the polarization rotation angle $\theta$ obtained through a 2D raster scan of a selected region of the sample. The first column shows the classical measurement acquired with a CMOS camera and a polarizer using the Malus' law $I=I_0\cos^22\theta$. The second column shows the quantum measurement obtained from Eq.~\ref{eq:angle_equation}. The top row presents a wide-area scan of the sample, and the bottom row shows a close-up of the region highlighted by the red square. We measured each point only once during $T=5.25$~s.}
\label{fig:HOM_sample}
\end{figure}

The experimental setup was then tested in the imaging configuration by raster-scanning an artificial sample consisting of a fused-silica slide onto which shards of a polymer retarder film with a designed wavelength at $560~nm$ of thickness $d_{HWP}\approx60\,\mu\text{m}$ were placed. As an initial step, we measured the HOM dip as a function of the delay of the reference-arm, $\Delta z$, for different positions of the sample under analysis. Regions with varying numbers of superimposed polymer retarder film shards were selected in order to investigate how the dip position depends on the effective sample thickness, as shown in Figure \ref{fig:HOM_vs_number_of_HWP}. The results clearly indicate that increasing the number of layers up to $2$ produces only a negligible shift in the dip position, where we measured a maximum shift of the dip $\Delta_z\approx100~\mu m\ll l_c$. Indeed, the delay introduced by a single layer is given by $d_{HWP}\left(n_{HWP}-1\right)\ll l_c$, where $n_{HWP}\sim1.5$ is the refractive index of the half-wave plate material. Thus, several shards can be stacked without significantly modifying the dip position. Having confirmed that the effect of sample thickness can be decoupled from that of polarization rotation, we then proceeded to image a square region of the sample and construct a birefringence map, shown in Figure \ref{fig:HOM_sample}. As a reference, we considered the image obtained with a single camera with a polarizer (analyzer) in front of it, obtaining a direct measurement of the rotation angle according to the Malus's law. Overall, we observe a good agreement between the two images, with enhanced boundary contrast in the visibility map, consistent with possible mode mismatch (e.g., scattering/aberration) near shard edges, which reduces interference visibility. All in all, these results confirm that operating at the broadened HOM-dip center of a narrowband photon pair source yields a thickness-insensitive visibility contrast that reproduces the expected polarimetric response while enhancing boundary sensitivity in the birefringence map.

\section{Conclusions}

The present study demonstrates a proof-of-concept for quantum birefringence imaging by adapting a Hong-Ou-Mandel (HOM) interferometer to perform a raster-scan microscopy. In particular, we exploit the use of a narrowband source to broaden the HOM dip, making the visibility insensitive to thickness-induced path-delay variations up to hundreds of micrometers, improving the robustness of polarization-based contrast. 

On the theory and estimation side, we established a complete statistical detection model (including loss) and derived a maximum-likelihood estimator for the polarization rotation angle. We further used Fisher-information analysis to identify the optimal operating point at the dip center and experimentally verified operation in a maximum-precision regime. In particular, it was demonstrated the saturation of the Cramér–Rao bound, meaning that given the number of recorded detection events, the estimator attains the minimum achievable variance allowed by the information content of the HOM measurement. 

On the imaging side, raster scans of polymer retarder-film shards produced spatial maps of HOM-dip visibility and corresponding $\theta$ estimates whose contrast followed the local optical-axis orientation, enabling inference of birefringence structure without losing the HOM effect due to thickness variation (e.g. superposition of distinct shards). Besides the overall agreement with the classical polarizer-based reference, the HOM-visibility maps showed improved delineation of shard boundaries, consistent with additional spatial-mode distinguishability (e.g., scattering/aberration) near edges that selectively degrades two-photon interference and therefore appears as enhanced boundary contrast.

Put into a broader perspective, the motivation for deploying quantum technologies in imaging is to realize quantum advantages, i.e. concrete performance benefits that mitigate (or surpass) practical and fundamental limitations of classical, intensity-based imaging. In the specific implementation reported here, three advantages are particularly direct. First, because birefringence information is inferred from two-photon coincidences (with a $1$~ns coincidence window), uncorrelated background light and detector dark counts contribute predominantly to accidental coincidences and are therefore strongly suppressed relative to the correlated-pair signal, which supports operation at low photon flux and in noisy environments. Second, the broadened dip (enabled by the narrowband source) relaxes path-length stability requirements: small axial misalignments and thickness variations that would significantly wash out interference in ultrabroadband HOM microscopy become comparatively negligible here, allowing polarization contrast to be extracted without nanometre-scale delay tracking at each pixel. Third, the HOM visibility provides an additional contrast channel that is intrinsically sensitive to any mechanism that introduces two-photon distinguishability (including spatial-mode mismatch), which can be leveraged as a feature, e.g., for edge-/defect-enhanced mapping in scattering or structured materials, rather than as a nuisance. 

Thus, and like other quantum imaging modalities, the combination of coincidence detection, robustness to uncorrelated noise, and low-illumination operation makes this approach a strong candidate for quantitative characterization of birefringent structures in photosensitive and biologically relevant samples where damage, background, or stability constraints are limiting. Yet, significant challenges remain to be tackled, including a two-fold ambiguity in $\theta$. Future work may progress on the implementation of dual-reference measurements to extend the estimator to jointly recover $\theta$ and retardance $\delta$, and evaluate performance in controlled scattering phantoms. Additionally, adaptive wavefront correction may also be explored to preserve interference visibility in highly scattering environments.

\begin{acknowledgments}
This work is co-financed by Component 5 - Capitalization and Business Innovation, integrated in the Resilience Dimension of the Recovery and Resilience Plan within the scope of the Recovery and Resilience Mechanism (MRR) of the European Union (EU), framed in the Next Generation EU, for the period 2021 - 2026, within project HfPT, with reference 41, and by national funds through FCT – Fundação para a Ciência e a Tecnologia, I.P., under the support UID/50014/2023 (10.54499/UID/50014/2023). Nuno A. Silva acknowledges the support of FCT under the grant 2022.08078.CEECIND/CP1740/CT0002 (10.54499/2022.08078.CEECIND/CP1740/CT0002). 
\end{acknowledgments}

\bibliographystyle{apsrev4-1}
\bibliography{quantumimagingbire}

\end{document}